\documentclass[aps,prb,twocolumn,showpacs]{revtex4}

\begin{document}
\title{Electromagnetic response of unconventional superconductors}

\author{V. P. Mineev}

\affiliation{Commissariat \`a l'Energie Atomique, INAC / SPSMS, 38054
Grenoble, France}
\date{\today}

\begin{abstract}
We derive the current response to the linearly polarized  electromagnetic field with finite frequency and wave vector incident normally
on the  specular surface of a clean nonconventional superconductor 
with orbital spontaneous magnetization parallel to the crystal axis and perpendicular to the crystal surface. 
The result includes the usual part known from the theory of conventional superconductivity  
and as well the magneto-optical term typical for the superconductors with spontaneous time reversal breaking. As an application of the basic current-field relation we consider the Kerr effect for the rotation of polarization of  infrared light reflected from the superconductor surface. 
\end{abstract}


\maketitle

\section{Introduction}
The theoretical investigation of the electromagnetic absorption in anisotropic superconductors has been put forward about two decades ago \cite{Klemm88,Hir89,Hir92}. The treatments were based on the expression for the current response to an electromagnetic vector potential well known for the conventional superconductors \cite{StPhys,FizKin}. The pairing state in nonconventional superconductors brings
in electrodynamic response its own anisotropy additional to the normal metal crystalline anisotropy.
Also, a nonconventional pairing state is suppressed by impurities and the superconductor surface.
These specific features of nonconventional superconductors are important as well for the collective mode excitation. And this circle of problems has been formulated and partly solved in the cited above papers. 

The lowering crystal symmetry in non-$s$-wave superconducting state could give rise to the changes of 
optical properties.
This subject based again on the expression for the current response to an electromagnetic vector potential for the conventional superconductors has been investigated in the paper \cite{Li} in application to the mixed $s+d$ and $s+id$ singlet states and to the mixed singlet-triplet state in the metal with broken both space and time reversal  symmetry. The latter superconducting state exhibits circular dichroism.
In the case of violation only the time reversal symmetry this phenomenon was found absent \cite{Li}.
More precisely, it is was shown, that for the effect to exist, that in addition to broken time-reversal symmetry, it is necessary to take into account the weak particle-hole assymmetry of the low energy excitations of the metallic state \cite{Yip}. The circular dichroism arises from the order parameter collective mode response of the superconductor.

In the present publication we shall show that in all the treatments mentioned above there was overlooked some specific term in the current response to the e-m field, which  exists even in the case of perfect particle-hole symmetry. This term is crucial for the magneto-optical phenomena in the superconducting states  
like A-phase of $^{3}He$ or  $E_{1g}$ state in hexagonal or tetragonal metals possessing of Cooper pair orbital ferromagnetism breaking the time reversal symmetry \cite{Sam}. 
The omission of this part of the current response had an academic character until the recent time when the observation of the Kerr effect that is rotation of polarization of reflected light from the surface of superconducting $Sr_2RuO_4$ has been reported \cite{Xia}. The theoretical treatments of this phenomenon has been proposed in \cite{Yak} by means a calculation of the corresponding part of the 
system effective action and in \cite{Min} in frame of two-fluid model using the 
results \cite{Muz,Bal} known from the theory of the superfluid $^3He-A$. Both of these approaches
present in principle the correct treatment of the Kerr effect for the superconductors with time reversal breaking. However, it is of interest to derive a general expression for the electromagnetic response 
(Matsubara susceptibility at arbitrary ${\bf q}$ and $\omega$) of the  nonconventional superconductor 
including both the usual part known from the theory of conventional superconductivity  \cite{FizKin}
and as well the new term typical for the superconductors with spontaneous time reversal breaking.
This derivation is given in the present publication. 

In view of complexity of the problem it is appropriate to mention  here the limitations of our treatment. We shall consider the linearly polarized uniform in space light normally incident on the surface of impurity free superconductor occupying the half of space $z<0$. We shall discuss only so called equal spin $p$-pairing state (A-phase)
possessing by the spontaneous magnetization due to the Cooper pair orbital momentum that we assume to be  perpendicular to the superconductor surface with specular boundary conditions for the quasiparticle reflection.   The singlet $d$-wave $E_{1g}$ pairing state $\Delta({\bf k})\propto \hat k_z({\hat k_x+i\hat k_y})$ as well as any superconducting state in an uniaxial crystal with orbital spontaneous magnetization parallel to the crystal axis and perpendicular to the crystal surface can be considered in the same manner. 
More general situation as well as the other particular cases will be considered elsewhere.

The obtained general current-field relation is applied to the finding of the Kerr rotation. The calculation is performed at high enough frequencies where taking into account the time dispersion one can neglect by the space dispersion and by the quasiparticle collisions with impurities. This frequency interval is
known as the infrared skin-effect region \cite{FizKin}.

\section{Matsubara susceptibility}
 
 As it is well known when a transverse electromagnetic field such that $div{\bf E}=0$ incident normally on the surface of isotropic superconductor with $s$-pairing the current-field relation  \cite{StPhys,FizKin}
has the simple form
\begin{equation}
{\bf j}(\omega,{\bf q})= -Q(\omega,{\bf q}){\bf A}(\omega,{\bf q}),
\end{equation}
where  $Q(\omega,{\bf q})$ is the Matsubara susceptibility and the ${\bf A}(\omega,{\bf q})$ is the transversal part of the vector-potential ${\bf q}{\bf A}(\omega,{\bf q})=0$. This case 
${\bf E}(\omega,{\bf q})=i\omega{\bf A}(\omega,{\bf q})$ and the scalar potential can be taken equal to zero $\varphi(\omega,{\bf q})=0$. We put velocity of light and the Planck constant $c=\hbar=1$ throughout the paper. On the other hand, as it follows from the Maxwell equation $rot{\bf H}=\varepsilon \partial{\bf E}/\partial t+4\pi{\bf j} $ there is no current divergency
$div{\bf j}=0$ and hence $\partial\rho/\partial t =0$. So, this case the collective plasma modes are absent. In application to the superconducting state this means that one can search the response of the current to the electromagnetic field not taking into consideration the spacial changes of the order parameter. Indeed, the straightforward calculation based on the Gor'kov equations and the self-consistency equation results in
the first order corrections to the $\Delta$ proportional to $div {\bf A}$ that is equal to zero.  

We consider linearly polarized purely transverse electromagnetic field normally (along the $z$-direction) incident from the vacuum to the boundary of an A-phase like superconductor \cite{Sam} characterized by the equal-spin pairing and the orbital part of the order parameter
\begin{equation}
\Delta({\bf k},{\bf r})=\Delta_i({\bf r})\hat k_i
\label{eA}
\end{equation}
In the equilibrium the order parameter is uniform $\Delta_i=\Delta(\hat x_i+i\hat y_i)$ and
\begin{equation}
\Delta({\bf k})=\Delta(\hat k_x+i\hat k_y)
\label{eB}
\end{equation}
 
 Under the influence of electromagnetic radiation the order parameter requires the coordinate dependent  amplitudes $\Delta_i({\bf r})$. However,  for the flat geometry, order parameter orientation and the light polarization described above the situation coincides exactly with the case of ordinary superconductivity (see the Ref.1,2,3). It means: one can 
work with  the coordinate independent amplitudes $\Delta_i=\Delta(\hat x_i+i\hat y_i)$, put the scalar electric potential equal to zero and ignore a collective modes excitations. Still, it does not mean that the order parameter completely given by its equilibrium shape (\ref{eB}).  The reason for this is its ${\bf k}$-dependence. Due to this in transverse electromagnetic field the argument of the  order parameter in the Gor'kov equations is shifted \cite{Bal}
$\Delta({\bf k}-e{\bf A}(\omega_n,{\bf q}))$.
Thus, in complete analogy with derivation given in the textbook \cite{FizKin}, we obtain for the 
the first order corrections  in ${\bf A}(\omega_n,{\bf q})$ to the Matsubara Green functions
\begin{eqnarray}
G^{(1)}(\tau_1, {\bf r}_1; \tau_2, {\bf r}_2)\nonumber\\
=G(\tau_1, {\bf r}_1; \tau_2, {\bf r}_2)-
G^{(0)}(\tau_1-\tau_2, {\bf r}_1-{\bf r}_2)\nonumber\\
=g(\tau_1-\tau_2, {\bf r}_1-{\bf r}_2)\exp\left[ 
\frac{i}{2}{\bf q}({\bf r}+{\bf r}')-\frac{i}{2}\omega_n(\tau+\tau')\right],
\label{e4}
\end{eqnarray}
\begin{eqnarray}
F^{+(1)}(\tau_1, {\bf r}_1; \tau_2, {\bf r}_2)\nonumber\\=F^+(\tau_1, {\bf r}_1; \tau_2, {\bf r}_2)-
F^{+(0)}(\tau_1-\tau_2, {\bf r}_1-{\bf r}_2)\nonumber\\
=f(\tau_1-\tau_2, {\bf r}_1-{\bf r}_2)\exp\left[ 
\frac{i}{2}{\bf q}({\bf r}+{\bf r}')-\frac{i}{2}\omega_n(\tau+\tau')\right],
\label{e5}
\end{eqnarray} 
the system of the algebraic Gor'kov equations
\begin{eqnarray}
\left[ i\left ( \Omega_m+\frac{\omega_n}{2}\right)-\varepsilon\left ( {\bf k}+\frac{{\bf q}}{2}\right)
+\mu\right]g(\Omega_m, {\bf q})\nonumber\\
+\Delta({\bf k})f(\Omega_m, {\bf q})\nonumber\\
=-e{\bf A}(\omega_n,{\bf q})\left[\frac{\partial\varepsilon({\bf k})}{\partial {\bf k}}G^{(0)}\left(\Omega_m-\frac{\omega_n}{2},
{\bf k}-\frac{{\bf q}}{2}\right)\right.\nonumber\\
\left.+\frac{\partial \Delta({\bf k})}{\partial {\bf k}}
F^{+(0)}\left(\Omega_m-\frac{\omega_n}{2},
{\bf k}-\frac{{\bf q}}{2}\right)
\right],
\label{e6}
\end{eqnarray} 
\begin{eqnarray}
\left[- i\left ( \Omega_m+\frac{\omega_n}{2}\right)-\varepsilon\left ( {\bf k}+\frac{{\bf q}}{2}\right)
+\mu\right]f(\Omega_m, {\bf q})\nonumber\\
-\Delta^*({\bf k})g(\Omega_m, {\bf q})\nonumber\\
=e{\bf A}(\omega_n,{\bf q})\left[\frac{\partial\varepsilon({\bf k})}{\partial {\bf k}}F^{+(0)}\left(\Omega_m-\frac{\omega_n}{2},
{\bf k}-\frac{{\bf q}}{2}\right)\right.\nonumber\\
\left.+\frac{\partial \Delta^*({\bf k})}{\partial {\bf k}}
G^{(0)}\left(\Omega_m-\frac{\omega_n}{2},
{\bf k}-\frac{{\bf q}}{2}\right)
\right].
\label{e7}
\end{eqnarray} 
Here, the
functions $g(\Omega_m, {\bf k})$ and $g(\Omega_m, {\bf k})$ are determined by the Fourier transformation 
 \begin{equation}
 g(\tau,{\bf r})=T\sum_{m=-\infty}^{\infty}\int g(\Omega_m, {\bf k})\exp[i{\bf k}{\bf r}-i\Omega_m\tau]
 \frac{d^3k}{(2\pi)^3},
 \end{equation}
 \begin{equation}
 f(\tau,{\bf r})=T\sum_{m=-\infty}^{\infty}\int f(\Omega_m, {\bf k})\exp[i{\bf k}{\bf r}-i\Omega_m\tau],
 \frac{d^3k}{(2\pi)^3},
 \end{equation}
and  the "nonperturbed" Green functions are given by
\begin{equation}
G^{(0)}(\omega,{\bf k})=-\frac{i\omega+\xi}{\omega^2+E^2},~~~
F^{+(0)}(\omega,{\bf k})=\frac{\Delta^*({\bf k})}{\omega^2+E^2},
\end{equation} 
\begin{equation}
E=\sqrt{\xi({\bf k})^2+|\Delta({\bf k})|^2},~~~~~\xi({\bf k})=\varepsilon({\bf k})-\mu,
\end{equation}
and the Matsubara frequencies are $\omega_n=2\pi nT$,  and $\Omega_m=(2m+1)\pi T$.

The solution of the system (\ref{e6}), (\ref{e7}) is 
\begin{equation}
g(\Omega_m, {\bf k})=g_1(\Omega_m, {\bf k})+g_2(\Omega_m, {\bf k}),
\end{equation}
\begin{eqnarray}
g_1(\Omega_m, {\bf k})=-e{\bf v}({\bf k}){\bf A}(\omega_n,{\bf q})\{ G^{(0)}(K_+)G^{(0)}(K_-)
\nonumber\\
+F^{(0)}(K_+)F^{+(0)}(K_-)\},
\end{eqnarray}
\begin{eqnarray}
g_2(\Omega_m, {\bf k})=-e{\bf A}(\omega_n,{\bf q})\{ G^{(0)}(K_+)
\frac{\partial \Delta({\bf k})}{\partial {\bf k}}F^{+(0)}(K_-)
\nonumber\\
+
F^{(0)}(K_+)\frac{\partial \Delta^*({\bf k})}{\partial {\bf k}}G^{(0)}(K_-)\}~~
\end{eqnarray}
Here for the brevity we have introduced notations
\begin{equation}
K_{\pm}=\left(\Omega_m\pm\frac{\omega_n}{2}, {\bf k}\pm\frac{{\bf q}}{2}\right),
\end{equation}
and ${\bf v}({\bf k})=\partial\varepsilon ( {\bf k})/\partial {\bf k}$.
As result the current density also consists of two terms
\begin{equation}
{\bf j}(\omega_n,{\bf q})={\bf j}_1(\omega_n,{\bf q})+{\bf j}_2(\omega_n,{\bf q}),
\end{equation}
where the first one is the standart current density known from the theory of conventional 
superconductivity \cite{FizKin}
\begin{eqnarray}
{\bf j}_{1}(\omega_n,{\bf q})=2eT\sum_{m=-\infty}^{\infty}\int {\bf v}({\bf k})g_1(\Omega_m, {\bf k})
 \frac{d^3k}{(2\pi)^3}\nonumber\\
 -\frac{Ne^2}{m^*}{\bf A}(\omega_n,{\bf q}),
\end{eqnarray}
here $m^*$ is the basal plane effective mass,
and the second one is the new term responsible for the magneto-optical phenomena
\begin{equation}
{\bf j}_2(\omega_n,{\bf q})=2eT\sum_{m=-\infty}^{\infty}\int {\bf v}({\bf k})g_2(\Omega_m, {\bf k})
 \frac{d^3k}{(2\pi)^3}.
 \end{equation}
These expressions can be rewritten in terms of the Matsubara's susceptibilities
\begin{equation}
{\bf j}_1(\omega_n,{\bf q})= -Q_1(\omega_n,{\bf q}){\bf A}(\omega_n,{\bf q}),
 \end{equation}
\begin{equation}
j_{2i}(\omega_n,{\bf q})= -Q_{2ij}(\omega_n,{\bf q})A_j(\omega_n,{\bf q}),
 \end{equation}
where
\begin{eqnarray}
Q_1(\omega_n,{\bf q})= \frac{Ne^2}{m^*}~~~~~~~~~~~~~~~~~~~\nonumber\\
+e^2T\sum_{m=-\infty}^{\infty}\int (v_x^2({\bf k})+v_y^2({\bf k}))
\{ G^{(0)}(K_+)G^{(0)}(K_-)\nonumber\\ 
+F^{(0)}(K_+)F^{+(0)}(K_-)\}
\frac{d^3k}{(2\pi)^3}.
\label{ex}
\end{eqnarray}
For $Q_{2ij}(\omega_n,{\bf q})$ after simple calculations, taking into account eqn. (\ref{eB}),
we obtain
\begin{eqnarray}
Q_{2ij}(\omega_n,{\bf q})=-
\frac{2\omega_nm^*e^2T}{k_F^2}~~~~~~~~~~~~~~~~\nonumber\\
\sum_{m=-\infty}^{\infty}\int v_i({\bf k})(\hat z\times {\bf v}({\bf k}))_j
F^{(0)}(K_+)F^{+(0)}(K_-)
\frac{d^3k}{(2\pi)^3}.
\label{ey}
\end{eqnarray}
The further well known procedure \cite{FizKin} consists of analytical continuation of these expressions from the discrete set of Matsubara frequences into entire half-plane $\omega>0$. The result for 
$Q_1(\omega,{\bf q})$ is known and has been used in \cite{Klemm88,Hir89,Hir92,Li} and we shall not write it here. The similar calculation for the $Q_2(\omega,{\bf q})$ yields
\begin{eqnarray}
Q_{2ij}(\omega,{\bf q})=
\frac{m^*e^2\omega}{k_F^2}
\int_{-\infty}^{\infty}\frac{d\Omega}{2\pi}
\int  v_i({\bf k})(\hat z\times {\bf v}({\bf k}))_j\nonumber\\
\tanh\frac{\Omega}{2T}\{[F^{(0)R}_+(\Omega)-F^{(0)A}_+(\Omega)]F^{(0)A}_-(\Omega-\omega)
\nonumber\\
+[F^{(0)R}_-(\Omega)
-
F^{(0)A}_{-}
(\Omega)]
F^{(0)R}_+(\Omega+\omega)\}\frac{d^3k}{(2\pi)^3}.
\label{ez}
\end{eqnarray}
Here the arguments ${\bf k}\pm{\bf q}/2$ of the Green functions are substituted by the subscripts
$\pm$, the superscripts  $R$ and $A$ are related to the retarded and advanced Green functions correspondingly 
\begin{equation}
F^{(0)R,A}(\omega)=\frac{\Delta({\bf k})}{2E}\left[\frac{1}{\omega+E\pm i\gamma}-
\frac{1}{\omega-E\pm i\gamma}\right].
\label{e24}
\end{equation} 
It is obvious  that $Q_{2ij}=-Q_{2ji}$.

The expression (\ref{ez}) is the basic equation for the investigation of magneto-optical phenomena in nonconventional superconductors  possessing by the spontaneous magnetization due the Cooper pair orbital momentum.
Let us apply it now to the calculation of the Kerr rotation of the reflected light polarization. 

\section{Kerr effect}

We consider the infrared light normally incident to the flat superconducting surface.  This case $\omega\gg\Delta$ and the presence of the gap in the quasiparticle spectrum of superconductor is unimportant.
Then we assume that the e-m field penetrates into the metal on the length $\delta$ which is much larger than $v_F/\omega$ and smaller than the mean free path
\begin{equation}
\frac{v_F}{\omega}\ll\delta<l.
\end{equation}
This frequency interval is
known as the infrared skin-effect region \cite{FizKin}. The first part of the inequality provides a possibility to neglect of the space dispersion, that is  put ${\bf  q}=0$ in the Eqns. (\ref{ex})-(\ref{ez}). The second part 
means collision free condition.
The penetration depth is  frequency independent 
\begin{equation}
\delta=\sqrt{\frac{m^*c^2}{4\pi e^2 n_e}},
\end{equation}
and it is typically $\approx 10^{-5} cm$. 
The conductivity is pure imaginary and given by
\begin{equation}
\sigma=i\sigma_{xx}^{\prime\prime}=\frac{ie^2n_e}{m^*\omega},
\label{e27}
\end{equation}
where $n_e$ is the conducting electron density.
The latter properties were found in the textbook \cite{FizKin} by means of kinetic equation. They can be established as well from the general formula for 
$Q_1(\omega, {\bf q})$ in $\Delta=0$ and ${\bf q}=0$
limit.

The off-diagonal component of conductivity are determined by  the equation
\begin{eqnarray}
\sigma_{xy}=\frac{i}{\omega}Q_{2xy}(\omega,{\bf q}=0)=
\frac{i m^*e^2}{k_F^2}
\int_{-\infty}^{\infty}\frac{d\Omega}{2\pi }
\int  v_x^2({\bf k})\nonumber\\
\tanh\frac{\Omega}{2T}\{[F^{(0)R}(\Omega)-F^{(0)A}(\Omega)]F^{(0)A}(\Omega-\omega)
\nonumber\\
+[F^{(0)R}(\Omega)
-
F^{(0)A}
(\Omega)]
F^{(0)R}(\Omega+\omega)\}\frac{d^3k}{(2\pi)^3}.
\label{e23}
\end{eqnarray}
Let us substitute  here the expression (\ref{e24}) for the Green functions. Then after the integration over $\Omega$ we obtain 
\begin{eqnarray}
\sigma_{xy}=-\frac{e^2}{2m^*}\int \hat{ k}_x^2|\Delta({\bf k})|^2\frac{\tanh\frac{E}{2T}}{E^2}
~~~~~~~~~~~~~~\nonumber\\
\left[\frac{1}{2E+\omega+i\gamma}+\frac{1}{2E-\omega-i\gamma} \right]\frac{d^3k}{(2\pi)^3}.
\end{eqnarray}
So, the off-diagonal conductivity is a complex function of frequency
\begin{equation}
\sigma_{xy}=\sigma_{xy}^\prime+i\sigma_{xy}^{\prime\prime}
\end{equation}
At large frequency $\omega\gg\Delta\gg\gamma=v_F/2l$  
the real part of it is determined mainly by the the first term of the integrand \begin{eqnarray}
\sigma_{xy}^\prime=-\frac{e^2}{m^*}\int \frac{dS_{\hat k}}{(2\pi)^3v_F({\bf k})}\hat{ k}_x^2
|\Delta({\bf k})|^2\nonumber\\
\int_{\Delta({\bf k})}^\infty\frac{\tanh\frac{E}{2T}}{E(2E+\omega)}\frac{dE}{\sqrt{E^2-|\Delta({\bf k})|^2}}
\end{eqnarray}
and the imaginary part by the second one
\begin{equation}
\sigma_{xy}^{\prime\prime}=-\frac{2\pi e^2}{m^*\omega^2}\int \frac{dS_{\hat k}}{(2\pi)^3v_F({\bf k})}\hat{ k}_x^2
|\Delta({\bf k})|^2.
\end{equation}
The integration is now performed over the Fermi surface.

The Kerr angle in the case of small absorption is given by \cite{Ben}
\begin {equation}
\theta=\frac{4\pi\sigma_{xy}^{\prime\prime}}{n(n^2-1)},
\end{equation} 
where the real part of the refraction index is determined by the  imaginary part of diagonal conductivity
\begin {equation}
n=\sqrt{\varepsilon-\frac{4\pi\sigma_{xx}^{\prime\prime}}{\omega}}=
\sqrt{\varepsilon-\left(\frac{\omega_p}{\omega}\right)^2}
\end{equation} 
The latter equality is written taking into account the eqn. (\ref{e27}) and the plasma frequency definition
$\omega_p=\sqrt{4\pi n_e e^2/m^*}$. It implies the validity of the condition $\omega>\omega_p/\sqrt{\varepsilon}$. Here $\varepsilon$ is the dielectric susceptibility at the limit of the infinite frequency.

\section{Conclusion}

We have derived the current response to the e-m field in nonconventional superconductors. In addition to the 
usual term well known both for conventional and nonconventional superconductors there is also an additional part given by eqn.(\ref{ez})  responsible for magneto-optical phenomena in nonconventional superconductors  possessing by the spontaneous magnetization due the Cooper pair orbital momentum. This part of response is not vanishing even if the particle-hole symmetry is perfectly fulfilled.

As an application to the developed formalism we have calculated the Kerr angle rotation for the infrared linearly polarized light incident on the flat specular  superconductor surface. The answer coincides with the found in the paper \cite{Min}, where has been used the electromagnetic field gauge with nonzero scalar potential. That case, of course, we did not obtain the total current response to the e-m field,
because it should include also the term proportional to the vector potential. However, for the calculation of imaginary part of off-diagonal component of conductivity tensor $\sigma_{xy}^{\prime\prime}$
this difference is unimportant.

\section*{ACKNOWLEDGMENTS}

The author is indebted to V.Yakovenko, R.Lutchyn, L.Bulaevskii and I.Kolokolov for the valuable discussions.

\end{document}